\documentclass[dvipdfm,conference]{IEEEtran}
\IEEEoverridecommandlockouts

\usepackage{cite}
\usepackage{amsmath,amssymb,amsfonts}
\usepackage{algorithmic}

\usepackage{textcomp}
\def\BibTeX{{\rm B\kern-.05em{\sc i\kern-.025em b}\kern-.08em
    T\kern-.1667em\lower.7ex\hbox{E}\kern-.125emX}}
\usepackage{tabularx}
\usepackage[pdftex]{graphicx}  
\usepackage{graphicx}
\usepackage{wrapfig}

\usepackage{epstopdf}

\newcommand{\snr}{\small \mbox{SNR}}

\newcommand{\sinr}{\small \mbox{SINR}}

\newcommand{\bA}{\boldsymbol{A}}
\newcommand{\bB}{\boldsymbol{B}}

\newcommand{\bD}{\boldsymbol{D}}

\newcommand{\bH}{\boldsymbol{H}}
\newcommand{\bI}{\boldsymbol{I}}

\newcommand{\bR}{\boldsymbol{R}}
\newcommand{\bS}{\boldsymbol{S}}

\newcommand{\bU}{\boldsymbol{U}}

\newcommand{\bW}{\boldsymbol{W}}
\newcommand{\bX}{\boldsymbol{X}}

\newcommand{\br}{\boldsymbol{r}}
\newcommand{\bs}{\boldsymbol{s}}

\newcommand{\bv}{\boldsymbol{v}}
\newcommand{\bw}{\boldsymbol{w}}
\newcommand{\bx}{\boldsymbol{x}}
\newcommand{\by}{\boldsymbol{y}}
\newcommand{\bz}{\boldsymbol{z}}

\newcommand{\bzero}{\boldsymbol{0}}

\newcommand{\bVb}{{\tilde \bVb}}

\newcommand{\cCN}{\mathcal{CN}}

\newcommand{\Nftil}{{\tilde N}_f}
\newcommand{\sfft}{{\rm  SFFT}}
\newcommand{\isfft}{{\rm  SFFT}^{-1}}
\newcommand{\Htil}{\tilde{\bH}}

\begin{document}

\title{How is Time Frequency Space Modulation Related to Short Time Fourier Signaling?
\thanks{Sayeed's work was partly supported by the US NSF through grants \#1703389 and \#1629713.}}

\author{\IEEEauthorblockN{Akbar M. Sayeed}
\IEEEauthorblockA{\textit{Department of Electrical and Computer Engineering}\\
\textit{University of Wisconsin}\\
\textit{Madison, WI 53711}\\
 akbar.sayeed@wisc.edu}}

\maketitle

\begin{abstract}
We investigate the relationship between Orthogonal Time Frequency Space (OTFS) modulation and Orthogonal Short Time Fourier (OSTF) signaling. OTFS was recently proposed as a new scheme for high Doppler scenarios and builds on OSTF. We first show that the two schemes are unitarily equivalent in the digital domain. However, OSTF defines the analog-digital interface with the waveform domain. We then develop a critically sampled matrix-vector model for the two systems and consider linear minimum mean-squared error (MMSE) filtering at the receiver to suppress inter-symbol interference. Initial comparison of capacity and (uncoded) probability of error reveals a surprising observation: OTFS under-performs OSTF in capacity but over-performs in probability of error. This result can be attributed to characteristics of the channel matrices induced by the two systems. In particular, the diagonal entries of  OTFS matrix  exhibit nearly identical magnitude, whereas those of the OSTF matrix exhibit wild fluctuations induced by multipath randomness. It is observed that by simply replacing the unitary matrix, relating OTFS to OSTF, by an arbitrary unitary matrix results in performance nearly identical to OTFS. We then extend our analysis to orthogonal frequency division multiplexing (OFDM) and also consider a more extreme scenario of relatively large delay and Doppler spreads. Our results demonstrate the significance of using OSTF basis waveforms rather than sinusoidal ones in OFDM in highly dynamic environments, and also highlight the impact of the level of channel state information used at the receiver.   
\end{abstract}

\section{Introduction}
\label{sec:intro}
The OTFS modulation was recently proposed as a new scheme for high Doppler scenarios \cite{hadani:wcnc17, hadani:long18} and builds on OSTF modulation proposed in \cite{Liu:01a,Kozek:98a}. However, the performance analysis in \cite{hadani:wcnc17, hadani:long18} is only relative to OFDM; in particular, a direct comparison with OSTF is missing. Subsequent works that build on OTFS include \cite{zemen:pimrc18} which explores other orthogonal precoding schemes and \cite{fettweis:gcom18} which relates OTFS to Generalized Frequency Division Multiplexing (GFDM), which shares many charcteristics with OSTF. However, surprisingly, neither of these works seems to be aware of OSTF \cite{Liu:01a}. Thus, all the recent OTFS developments are missing a comparison with OSTF. The goal of this paper is to clarify the intimate relationship between OTFS and OSTF and to compare their performance in relation to OFDM and related schemes. 
\begin{figure}[htb]
\begin{tabular}{c}
\hspace{-3mm}
\centerline{\includegraphics[width=0.48\textwidth]{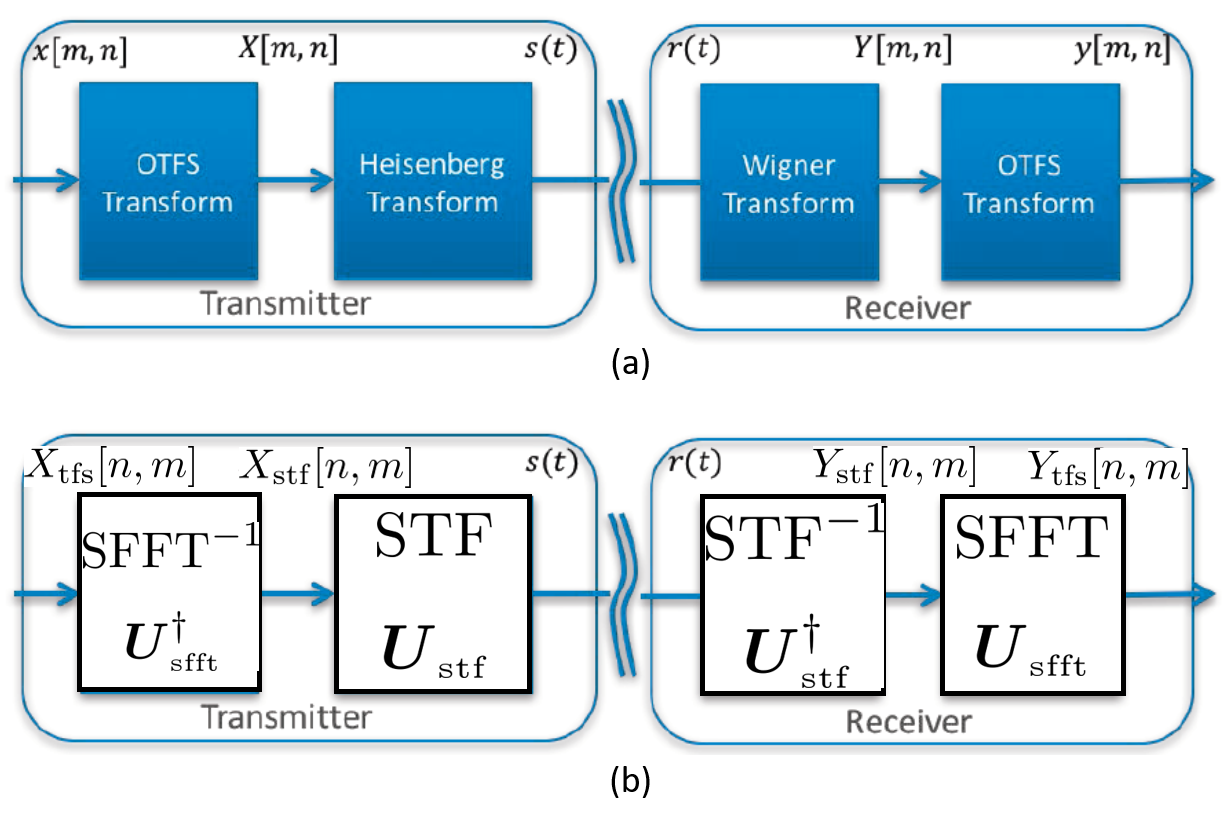}} 
\end{tabular}
\vspace{-4mm} 
\caption{\footnotesize{\sl Schematic showing the fundamental relationship between the OTFS and OSTF systems. (a) A schematic of the OTFS system from \cite{hadani:wcnc17}. (b) A schematic of the OTFS system in terms of OSTF modulation and SFFT.}}
\label{fig:otfs_ostf}
\vspace{-6mm} 
\end{figure}

We first describe the two systems in the context of communication over doubly dispersive channels in Sec.~\ref{sec:ltv} and show that the two modulation schemes are unitarily equivalent in the digital domain and thus share many fundamental characteristics. The fundamental relationship between the two systems is illustrated in Fig.~\ref{fig:otfs_ostf}. Essentially, the OTFS system is obtained by sandwiching the OSTF system between two digital unitary transforms, one at the transmitter (TX) and one at the receiver (RX). The transform is the Symplectic Finite Fourier Transform (SFFT) \cite{hadani:wcnc17} which is essentially a two-dimensional discrete Fourier transform (DFT) that operates on the matrix of time-frequency (or delay-Doppler) symbols. However, the OSTF modulation  defines the analog-digital interface.   We then develop a critically sampled $N \times N$ matrix-vector representation for both systems in Sec.~\ref{sec:matrix} and  then develop a system model in Sec.~\ref{sec:perf} to compare them and to analyze their performance. At the RX, we characterize the minimum-mean-squared-error (MMSE) filter matrix $\bW$ to suppress the interference between different time-frequency (OSTF) or delay-doppler (OTFS)  symbols. We then characterize the signal-to-interference-and-noise (SINR) for the $N$ channels, as a function of the operating signal to noise ratio (SNR), that in turn leads to analytical estimates of the channel capacity.  We present numerical results in Sec.~\ref{sec:disc}  to compare the capacity and (uncoded) probability of error, $P_e$, of OSTF, OTFS, and OFDM systems. The results for capacity and $P_e$ in a moderately dispersive channel lead to a surprising observation: {\em OTFS exhibits a  loss in capacity but a lower $P_e$ relative to OSTF.} Interestingly, the OFDM performance is between OSTF and OTFS. 

The reported results on $P_e$ and SINR confirm the remarkable {\em ``constant gain''} property of OTFS reported in \cite{hadani:wcnc17}. We further investigate this difference through the properties of the induced channel matrices in OTFS and OSTF. In particular, we know from \cite{Liu:01a} that OSTF basis waveforms serve as approximate eigenfunctions of doubly-selective channels and thus the OSTF channel matrix should be diagonally dominant. We introduce a {\em channel diagonality metric}, $\gamma$, to quantify this aspect and show that OTFS exhibits a significantly lower $\gamma$ which is related to its loss in capacity relative to OSTF.  To further investigate $P_e$ performance, we consider another system, called OSTF-U, that replaces the unitary SFFT precoding in OTFS with an arbitrary unitary matrix. Surprisingly,  OSTF-U exhibits capacity and $P_e$ performance that is nearly identical to OTFS. This is related to the result in \cite{zemen:pimrc18} that all constant-modulus unitary precoding schemes exhibit the ``constant gain'' property of OTFS.    

In a final set of results, we compare the performance of OTFS/OSTF-U, OSTF, and OFDM in a highly dispersive channel to explore the significance of appropriately chosing the underlying time-frequency basis waveforms defined by OSTF, and implicitly utilized in all recent works on OTFS, including \cite{hadani:wcnc17,hadani:long18,zemen:pimrc18,fettweis:gcom18}.  The results indicate that with full channel state information at the RX, the differences in performance of OTFS/OSTF-U, OSTF and OFDM get accentuated. On the other hand, if receiver design is only based on the diagonal entries of the channel matrix, as suggested by the eigen-property of OSTF, the  results show that OSTF performance is  significantly superior to OTFS/OSTF-U and OFDM, both in terms of capacity and $P_e$.

\vspace{-1mm}
\section{The OSTF and OTFS Systems for Doubly Dispersive Channels}
\label{sec:ltv}
The system model for communication over a doubly-dispersive multipath channel can be expressed as 
\begin{equation}
r(t) = \int H(t,f) S(f)e^{j2\pi f t} df + w(t)
\label{Htf_sys}
\end{equation}
where $S(f) = \int s(t) e^{-j2\pi f t} dt$  is the Fourier transform of the transmitted signal $s(t)$, $r(t)$ is the received signal, $w(t)$ is complex additive white Gaussian noise (AWGN),  and $H(t,f)$ is the time-varying frequency response of the channel 
\begin{equation}
H(t,f) = \sum_{\ell=1}^{N_p} \alpha_\ell e^{-j2\pi \tau_\ell f} e^{j2\pi \nu_\ell t}
\label{Htf}
\end{equation}
where $N_p$ denotes the number of paths, and $\alpha_\ell$, $\tau_\ell$ and $\nu_\ell$ represent the complex amplitude, delay, and Doppler shift, respectively, associated with the $\ell$-th path. The path delays and Doppler shifts lie within the channel spreads:
\begin{equation}
\tau_\ell \in [0, \tau_{max}] \ ; \ \nu_\ell \in [-\nu_{max}, \nu_{max}]
\label{ch_spreads}
\end{equation} 
where $\tau_{max}$ is the delay spread and $\nu_{max}$ the Doppler spread.

\subsection{The OSTF System}
\label{sec:ostf}
The transmitted OSTF signal is given by \cite{Liu:01a}
\begin{equation}
s(t) = \sum_{n = 0}^{N_t-1} \sum_{m=-\Nftil}^{\Nftil} X_{\rm stf} [n,m] g(t-n T_o) e^{j2\pi m F_o t}
\label{stf_mod}
\end{equation}
where $N = TW = N_t N_f$, $\Nftil = (N_f-1)/2$, where we assume that $N_f$ is odd, and $X_{\rm stf}[n,m]$ are the digital symbols to be modulated onto the OSTF basis waveforms given by 
\begin{align}
& \phi_{n,m}(t)  = g(t-nT_o)e^{j2\pi m F_o t}  \nonumber \\
& n = 0, 1, \cdots, N_t-1 \  ;  \  m = -\Nftil, \cdots, 0, \cdots, \Nftil
\label{stf_basis}
\end{align}
We assume that $g(t)$ is a unit energy rectangular pulse of duration $T_o$ resulting in an orthogonal STF basis. The (baseband) OSTF transmitted signal occupies a duration of $T = N_t T_o$ over a two-sided bandwidth $W=N_f F_o$.  The time and frequency shift parameters in (\ref{stf_mod}) satisfy $T_o F_o=1$. However, their relative values do have an impact on performance. The optimal choice, from the viewpoint of minimizing interference between the OSTF basis waveforms at the RX, is given by \cite{Liu:01a} 
\begin{equation}
\frac{T_o}{F_o} = \frac{\tau_{max}}{2\nu_{max}} \Longrightarrow T_o = \sqrt{\frac{\tau_{max}}{2\nu_{max}}} \ , \ F_o = \sqrt{\frac{2\nu_{max}}{\tau_{max}}}
\label{ToFo}
\end{equation}
At the RX, the OSTF demodulation is performed on $r(t)$ to compute the sufficient statistics for symbol detection
\begin{align}
Y_{\rm stf}[n,m]  & = \int_{0}^T r(t) g^*(t-nT_o)e^{-j2\pi m F_o t} dt 
\label{stf_demod}
\end{align}
The OSTF modulation in (\ref{stf_mod}) is referred to as the Heisenberg transform and the OSTF demodulation in (\ref{stf_demod}) as the Wigner transform in \cite{hadani:wcnc17}; see Fig.~\ref{fig:otfs_ostf}(a).

\subsection{The OTFS System}
\label{sec:otfs}
As noted in \cite{hadani:wcnc17, hadani:long18}, a key difference between OSTF and OTFS is that while the former modulates symbols in the time-frequency domain, as in (\ref{stf_mod}), the latter modulates symbols in the delay-Doppler domain. This is accomplished by performing $\isfft$ and $\sfft$ operations at the TX and RX, respectively, as depicted in Fig.~\ref{fig:otfs_ostf}(b).  Specifically, the transmitted Doppler-delay digital symbols in OTFS, denoted by $X_{\rm tfs}[k,\ell]$ are related to the time-frequency symbols $X_{\rm stf}[n,m]$ through $X_{\rm tfs}[k,\ell]  = \sfft\{ X_{\rm stf}[n,m] \}$ 
\begin{equation}
X_{\rm tfs}[k,\ell]  =  \frac{1}{\sqrt{N_t N_f}} \sum_{n=0}^{N_t -1} \sum_{m = 0}^{N_f -1} X_{\rm stf}[n,m] e^{-j\frac{2\pi k n}{N_t}} e^{j\frac{2\pi \ell m}{N_f}}   \label{sfft} 
\end{equation}
and $X_{\rm stf}[n,m]   = \isfft\{ X_{\rm tfs}[k,\ell] \}$
\begin{equation}
 X_{\rm stf}[n,m] =  \frac{1}{\sqrt{N_t N_f}} \sum_{k=0}^{N_t -1} \sum_{\ell = 0}^{N_f -1} X_{\rm tfs}[k,\ell]  e^{j\frac{2\pi k n}{N_t}} e^{-j\frac{2\pi \ell m}{N_f}}  \label{isfft} 
\end{equation}
Note that $\sfft$ and $\isfft$ are unitary transforms and, thus, as depicted in Fig.~\ref{fig:otfs_ostf}(b), the OTFS system can be viewed as the OSTF system, with additional pre-processing by $\isfft$ at the TX and post-processing by $\sfft$ at the RX.

\section{Matrix Representation of OTFS and OSTF}
\label{sec:matrix}
 It is instructive to develop a matrix-based system representations for OTFS and OSTF, dictated by sampling theory, to further analyze their relationship.  For a signaling duration $T$ and two-sided bandwidth $W$, the critical sampling in time (and delay) and frequency (and Doppler) is given by \cite{sayeed:handbook08}
\begin{equation}
\Delta t = \Delta \tau = \frac{1}{W}  \ ; \ \Delta f = \Delta \nu = \frac{1}{T} 
\label{tf_sampling}
\end{equation}  
This yields $N = T/\Delta t = TW$ samples in time, over the signaling duration $T$, and $N = W/\Delta f = TW$ samples in frequency over the bandwidth $W$. 
For example, defining the sampled temporal signal as $s[n] = s(n \Delta t)$ and its sampled Fourier transform as $S[k] = S(k \Delta f)$ we get the relationship
\begin{align}
S[k]   = S(k \Delta f) & = \frac{1}{\sqrt{N}} \sum_{n=0}^{N-1} s[n] e^{-j\frac{2\pi nk}{N}} \label{FT}   \\
& \Updownarrow \nonumber \\
\bS  = \bU_N^\dagger \bs \  ; \  \bs = \bU_N \bS \ & ; \ \bU_N[n,k] = \frac{1}{\sqrt{N}} e^{j\frac{2\pi nk}{N}} \label{Udft}   
\end{align}
where $\bS$ and $\bs$ are N-dimensional vectors, $\bU_N$ is an $N \times N$ unitary discrete Fourier transform (DFT) matrix, and the superscript $^\dagger$ denotes the complex conjugate transpose.  

The input-output relationship in (\ref{Htf_sys}) can be expressed as 
\begin{align}
& \br    = \Htil \bU_N^\dagger \bs  + \bw  \   ; \ \Htil[n,m]  = \bH[n,m]e^{j\frac{2\pi nm}{N}} \nonumber \\
 & \bH[n,m]  = H(n\Delta t, m \Delta f) \  ; \ \bw \sim \cCN(\bzero, \bI_N) \ . 
\label{tf_sys_mat}
\end{align}

\subsection{The OSTF System}
\label{sec:ostf_mat}
The OSTF modulation  in (\ref{stf_mod}) can be expressed  as
\begin{equation}
\bs = \bU_{\rm stf} \bx_{\rm stf}  \ ; \ \bx_{\rm stf} = \bU_{\rm stf}^\dagger \bs
\label{stf_mod_mat}
\end{equation}
where $\bx_{\rm stf}$ is an $N$-dimensional vector representing $X_{\rm stf}[n,m]$ and $\bU_{\rm stf}$ is an $N \times N$ matrix whose columns are sampled versions of the  OSTF basis functions in (\ref{stf_basis}). Combining (\ref{tf_sys_mat}), (\ref{stf_mod_mat}) and the sampled representation of the OSTF demodulation in (\ref{stf_demod}) we get the overall relationship between the input to the OSTF modulator and the output of the OSTF demodulator
\begin{align}
\by_{\rm stf}  &  =  \bU_{\rm stf}^\dagger \br  =  \bH_{\rm stf} \bx_{\rm stf} + \bw_{\rm stf} \nonumber \\
\bH_{\rm stf} & =  \bU_{\rm stf}^\dagger \Htil \bU_N^\dagger \bU_{\rm stf} \ ; \ \bw_{\rm stf} = \bU_{\rm stf}^\dagger \bw  
\label{stf_sys_mat}
\end{align}
where $\by_{\rm stf}$ is a vector representation of $Y_{\rm stf}[n,m]$ in Fig.~\ref{fig:otfs_ostf}(b).  

\subsection{The OTFS System}
\label{sec:otfs_mat}
To develop a matrix-vector representation for the OTFS system we need the corresponding representation for $\sfft$ and $\isfft$. Let $\bX_{\rm stf}$ represent the $N_t \times N_f$ matrix formed by $X_{\rm stf}[n,m]$ and $\bX_{\rm tfs}$ represent the $N_t \times N_f$ matrix formed by $X_{\rm tfs}[n,m]$  in Fig.~\ref{fig:otfs_ostf}(b). Then, the $\sfft$ and $\isfft$ relationships in (\ref{sfft}) and (\ref{isfft}) can be expressed as 
\begin{equation}
\bX_{\rm tfs} = \bU_{N_t}^\dagger \bX_{\rm stf} \bU_{N_f} \ ; \ \bX_{\rm stf} = \bU_{N_t} \bX_{\rm tfs} \bU_{N_f}^\dagger
\label{sfft_mat}
\end{equation}
where $\bU_{N_t}$ and $\bU_{N_f}$ are unitary DFT matrices of dimension $N_t$ and $N_f$ as  in (\ref{Udft}).  Vectorizing  (\ref{sfft_mat}) we get
\begin{align}
\bx_{\rm stf} = {\rm vec}(\bX_{\rm stf}) \ & ; \ \bx_{\rm tfs}  = {\rm vec}(\bX_{\rm tfs}) \nonumber \\
\bx_{\rm stf}  = \bU_{\rm sfft}^\dagger \bx_{\rm tfs} \ & ; \ \bx_{\rm tfs} = \bU_{\rm sfft} \bx_{\rm stf} \nonumber \\
  \bU_{\rm sfft}  = \bU_{N_f}^{\dagger *} \otimes \bU_{N_t}^\dagger \ & ; \ \bU_{\rm sfft}^{-1} = \bU_{\rm sfft}^\dagger \ .
\label{sfft_vec}
\end{align}
where $\otimes$ represents the kronecker product, ${\rm vec}(\cdot)$ represents the vectorizing operation, and we have used the relationship ${\rm vec}(\bA \bD \bB) = \left [ \bB^{\dagger *} \otimes \bA \right ] {\rm vec}(\bD)$ \cite{Brewer}.

\begin{figure}[htb]
\begin{tabular}{c}
\centerline{\includegraphics[width=0.4\textwidth]{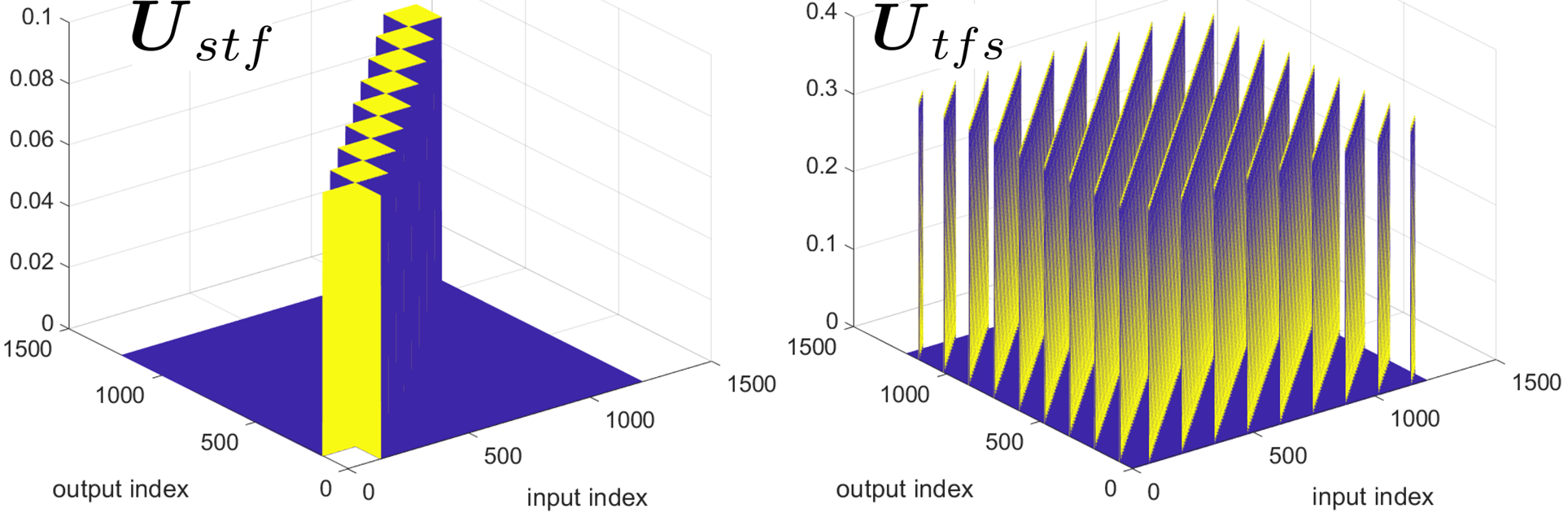}} 
\end{tabular}
\caption{\footnotesize{\sl Mesh plots for  $\bU_{\rm stf}$ and $\bU_{\rm tfs}$ for $N_t$ = $9$, $N_f$=$135$, $N$=$1215$}}
\label{fig:Ustf_tfs}
\end{figure}

Note that $\bU_{\rm sfft}$ is an $N \times N$ unitary matrix.   Per the definition in \cite{hadani:wcnc17}, the matrix representation for the OTFS modulation is 
\begin{equation}
\bU_{\rm tfs} = \bU_{\rm stf} \bU_{\rm sfft}^\dagger  
\label{Utfs}
\end{equation}
so that $\bs = \bU_{\rm tfs} \bx_{\rm tfs}$ and $\by_{\rm tfs} = \bU^\dagger_{\rm tfs} \br$ in Fig.~\ref{fig:otfs_ostf}(b);  $\by_{\rm tfs}$ is a vector representation of $Y_{\rm tfs}[n,m]$ in Fig.~\ref{fig:otfs_ostf}(b). Fig.~\ref{fig:Ustf_tfs} shows mesh plots for $\bU_{\rm stf}$ and $\bU_{\rm tfs}$. 
Combining (\ref{sfft_vec}) with (\ref{stf_sys_mat}) we get the overall relationship between the input to the OTFS modulator and the output of the OTFS demodulator
 \begin{align} 
\by_{\rm tfs}  &  =  \bU_{\rm tfs}^\dagger \br = \bH_{\rm tfs} \bx_{\rm tfs} + \bw_{\rm tfs} \nonumber \\
\bH_{\rm tfs} & =  \bU_{\rm sfft} \bH_{\rm stf} \bU_{\rm sfft}^\dagger \ ;  \  \bw_{\rm tfs} = \bU_{\rm tfs}^\dagger \bw  \ . 
\label{tfs_sys_mat}
\end{align}

\section{System Model for Analyzing OTFS and OSTF}
\label{sec:perf}
We now present a system model for analyzing and simulating the two systems and comparing their performance.  The input-output relationship between the input to the modulators at the TX and the output of the demodulators at the RX is  
\begin{equation}
\by= \sqrt{\snr} \bH \bx + \bw  
\label{sys_mod_io}
\end{equation}
where $\by$,  $\bH$, and $\bx$ represent either the OSTF system in (\ref{stf_sys_mat}) or the OTFS system in (\ref{tfs_sys_mat}) with the appropriate subscripts. The noise vector $\bw \sim \cCN(\bzero, \bI_N)$  and $\snr$ denotes the {\em average} SNR per dimension at the RX. 
\subsection{Interference Suppression at the Receiver}
\label{sec:mmse}
In general, both OSTF and OTFS will experience some interference (inter-channel and inter-symbol) at the RX between transmitted symbols. Thus, we consider linear minimum mean-squared error (MMSE) processing at the RX to suppress the interference. Let $\bW_{\rm stf}$ and $\bW_{\rm tfs}$ denote the $N \times N$ MMSE interference suppression matrices for the two systems. The outputs of the MMSE filter are
\begin{equation}
\bz_{\rm stf} = \bW_{\rm stf}^\dagger \by_{\rm stf} \ ; \ \bz_{\rm tfs} = \bW_{\rm tfs}^\dagger \by_{\rm tfs}  \ .
\label{mmse}
\end{equation}
We assume perfect CSI at the RX;  $\bH = \bH_{\rm stf}$ or $\bH = \bH_{\rm tfs}$ is known at the RX. The MMSE filter matrices are given by
\begin{equation}
\hspace{-0mm} \bW  \!=  \! \bR^{-1} \bH;  \ \bR \! = \! E[ \by \by^\dagger] = \snr \bH \bH^\dagger + \bR_{w} ; \ \bR_w \! = \! \bI_N   \label{W_mmse}   
 \end{equation}
where we have assumed independent unit-power digital symbols for both $\bx = \bx_{\rm stf}$ or $\bx = \bx_{\rm tfs}$; that is, $E [\bx\bx^\dagger] = \bI_N$, where $\bI_N$ denotes the identity matrix of dimension $N$.  The output of the MMSE filter in the two systems is given by 
\begin{align}
\bz  & = \bW^\dagger \by = \sqrt{\snr} \bH_{\rm c} \bx + \bv  \nonumber \\
 \bH_{\rm c} & = \bW^\dagger \bH \ , \ \bv \sim \cCN(\bzero, \bW^\dagger \bW)
\label{sys_mod_io_mmse}
\end{align}
where $\bH_{\rm c}$ represents the {\em composite} channel matrix that includes the MMSE filter. 

\vspace{1mm}
\noindent
{\bf Channel Diagonality Metric.} The differences in the performance of OSTF and OTFS are intimately related to the two key channel matrices: $\bH$ and $\bH_{\rm c} = \bW^\dagger \bH$.  The input and output dimensions correspond to the symbols transmitted and received in an uncoded system.  The diagonal terms reflect the channel strength coupling the corresponding input-output dimensions and are thus a measure of SNR in the absence of interference. The off-diagonal terms  represent the interference between the different input-output dimensions. Since OSTF basis functions serve as approximate eigenfunctions of the channel  \cite{Liu:01a}, we expect the matrices to be more diagonally dominant  for OSTF compared to OTFS. To explore this further, we define a {\em channel diagonality metric} $\gamma$:
 \begin{equation}
\gamma(\bH) = \frac{\sum_{n=1}^N |\bH[n,n]|^2}{\sum_{n=1}^N \sum_{m=1}^N |\bH[n,m]|^2}
\label{gamma}
\end{equation}
 which is the ratio of the  power in the diagonal entries to the total  power. The diagonal terms of $\bH_{\rm stf}$ are approximately given by $H(t,f)|_{t=nT_o, f = mF_o}$ \cite{Liu:01a}.

\subsection{SINR and Capacity Estimates}
\label{sec:cap}
 The SINR in the different dimensions of the MMSE filter output can  be calculated as
\begin{equation}
\sinr(n) = \frac{\snr |\bH_{\rm c}(n,n)|^2}{\sum_{i=1, i\neq n}^N \snr |\bH_{\rm c}(n,i)|^2 + \bR_v(n,n)}  \
\label{sinr}
\end{equation}
where $n=1, \cdots, N$ and $\bR_v = E[\bv \bv^H] = \bW^\dagger \bW$. 
Using the SINR values,  the capacity per dimension, conditioned on a  particular channel realization, can be estimated as 
\begin{equation}
C(\bH) = \frac{1}{N}\sum_{n=1}^N \log_2(1 + \sinr(n)) \ (\mbox{bits/sec/Hz - bps/Hz}) 
\label{C_H}
\end{equation} 
where we have assumed equal power transmission in each dimension and we note that the SINRs depend on $\bH$. The ergodic capacity {\em per dimension}, averaged over realizations of $\bH$, is  $C = E_{\bH}[C(\bH)]$, and the total capacity is $C \times W$ (bps). 

\vspace{1mm}

\noindent
{\bf Related Systems: OFDM, OSTF-U, and EIG.}
The differences between OSTF and OTFS are also related to the nature of the underlying modulation. As shown in \cite{Liu:01a}, the OSTF basis functions, which are time- and frequency-shifted versions of a time-frequency localized pulse, serve as approximate eigenfunctions for (underspread) doubly dispersive channels, as an extension of the sinusoidal basis functions in OFDM that serve as eigenfunctions for time-invariant, frequency-selective channels. The OFDM system is a special case of the OSTF system with $N_t = 1 \ (T_o = T)$ and $N_f = N \ (F_o = W/N)$  in (\ref{stf_mod}), appropriate for non time-selective channels \cite{Liu:01a}. The preprocessing with $\isfft$, on the other hand, makes OTFS analogous to spread-spectrum signaling since the delay-Doppler symbols are now spread over all the OSTF basis functions, as also noted in \cite{hadani:long18}.
So, a natural question is: {\em What kind of performance in induced by replacing SFFT with a different unitary transform?}
We thus also consider an OSTF-U system that uses an arbitrary unitary transformation $\bU$ as opposed to $\bU_{\rm sfft}$ in OTFS (see (\ref{Utfs})); that is, $\bU_{\rm stf-u}$ = $\bU_{\rm stf} \bU$.  In Sec.~\ref{sec:disc}, $\bU$ was generated by orthogonalizing the columns of a  random $N \times N$ matrix with $\cCN(0,1)$ entries. Using a DFT matrix $\bU = \bU_N$ also yields identical performance; not shown.  As a benchmark we also consider the EIG system that communicates over the singular vectors of $\bH$ (no interference between the singular vectors).  Let $\{\lambda_n\}$ denote the eigenvalues of $\bH^\dagger \bH$,  which are  identical for both $\bH_{\rm stf}$ and $\bH_{\rm tfs}$ since they are unitarily equivalent. The capacity of the EIG system, with equal power allocation in each dimension, is given by
\begin{equation}
C_{eig}(\bH) = \frac{1}{N} \sum_{n=1}^N \log_2(1 + \snr \lambda_n) \ (\mbox{bps/Hz}) \ .
\label{Ceig_H} 
\end{equation}
Note that the EIG system serves as a benchmark  and requires channel state information (CSI) at both the TX and RX.
  
\section{Numerical Results and Discussion}
\label{sec:disc}
The OTFS and OSTF systems are unitarily equivalent; see Fig.~\ref{fig:otfs_ostf}.  The OTFS system builds on the OSTF system via unitary pre-processing at the TX with $\isfft$ and unitary post-processing at the RX with $\sfft$.  We further analyze the relationship between OTFS, OSTF, OFDM and OSTF-U and illustrate with numerical results. For simulation, we consider system parameters similar to those in \cite{hadani:wcnc17}. 
\begin{figure}[hbt]
\vspace{-2mm}
\centerline{\includegraphics[width=3.1in]{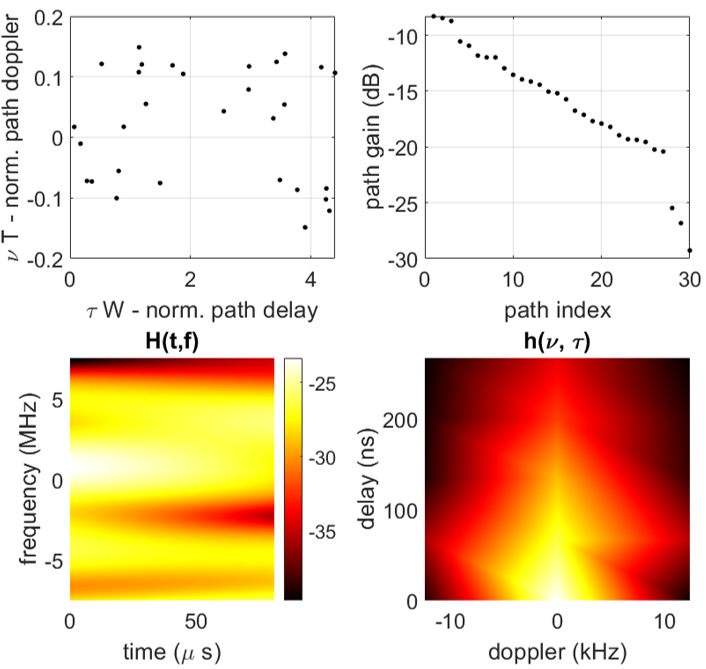}}
\vspace{-3mm}
\caption{\footnotesize{\sl Plots of the multipath parameters and images of $H(t,f)$ and $h(\nu,\tau)$.}}
\label{fig:mpc_H}
\end{figure} 
We consider a system operating at a carrier frequency of $f_c = 4$ GHz, with bandwidth $W=15$ MHz operating over a time-varying multipath channel with $\tau_{max} = 300$ns and $\nu_{max}=1.85$ kHz (corresponding to a maximum speed of 500 kmph). We simulate a channel with $N_p = 30$ paths with $\tau_\ell \sim {\rm unif}[0,\tau_{max}]$, $\nu_\ell \sim {\rm unif}[-\nu_{max},\nu_{max}]$ and $\alpha_\ell \sim \cCN(0,\sigma^2_\ell)$, with exponentially decreasing path powers $\sigma_\ell^2 = e^{-\ell/Np}$, resulting in a 4.2dB difference between the highest and lowest path powers. We also normalize the path powers so that $\sum_\ell \sigma^2_\ell = 1$, which is proportional to the average channel power used to define the average SNR in (\ref{sys_mod_io}). The OSTF system is designed using (\ref{ToFo}) resulting in $T_o  = 9 \mu$s and $F_o = 111.11$ kHz. The dimension of the OSTF system is $N=1215$, with $N_t=9$ and $N_f=135$. The OSTF symbol duration is $T=N_tT_o = 81 \mu$s and $W=N_f F_o = 15$MHz. Fig.~\ref{fig:mpc_H} plots the normalized path delays,  Doppler shifts,  powers, and images of $H(t,f)$  (and its 2D Fourier transform $h(\nu,\tau)$ - the delay-Doppler spreading function) for a particular realization.   
\begin{figure}[htb]
\vspace{-5mm}
\centerline{\includegraphics[width=3.2in]{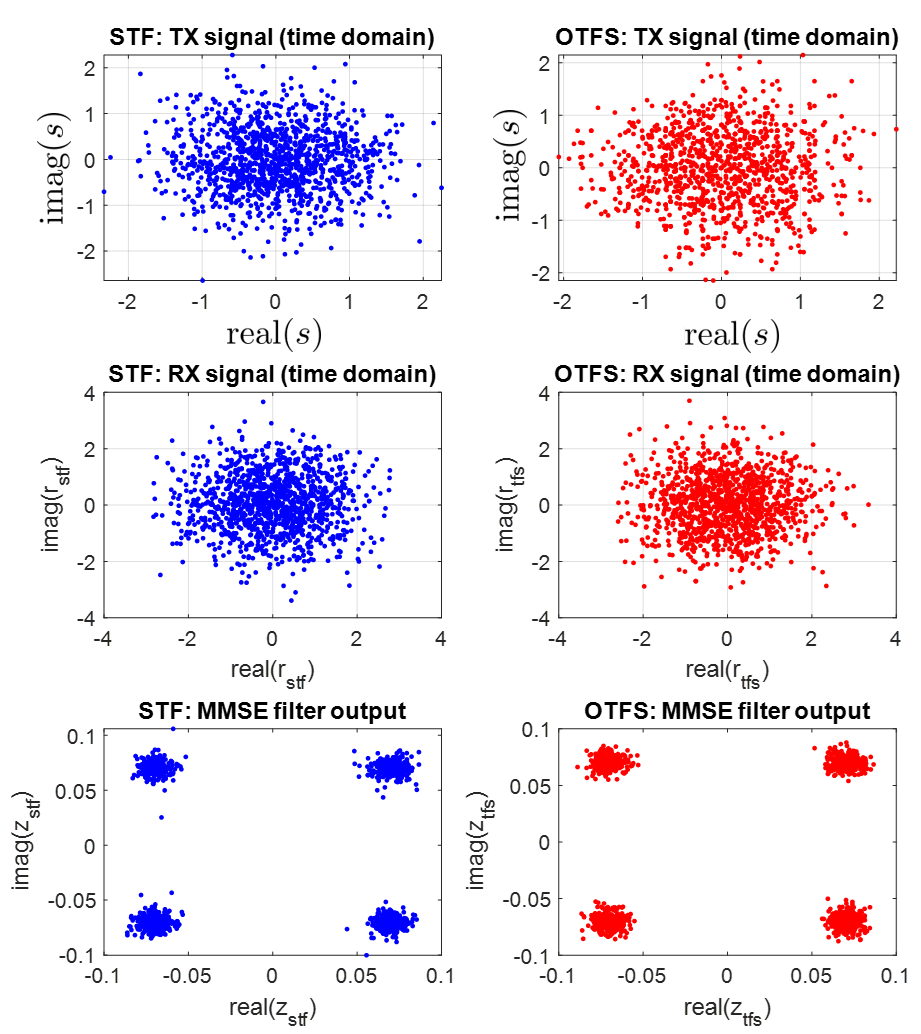}}
\vspace{-2mm}
\caption{\footnotesize{\sl Plots of the TX signal, RX signal, and the outputs of the MMSE filter for the OSTF and OTFS systems.}}
\label{fig:x_r_z_sigs}
\vspace{-3mm}
\end{figure} 

We consider 4-QAM constellation and Fig.~\ref{fig:x_r_z_sigs} plots the transmitted signal vector $\bs$, the received signal $\br$, and the outputs of the MMSE filters $\bz$ for a single transmission of an OSTF and OTFS packet (for the same 4-QAM input symbols $\bx_{\rm stf}=\bx_{\rm tfs}$) at an SNR per dimension of $20$dB. 
\begin{figure}[bht]
\vspace{-0mm}
\centerline{\includegraphics[width=3.1in]{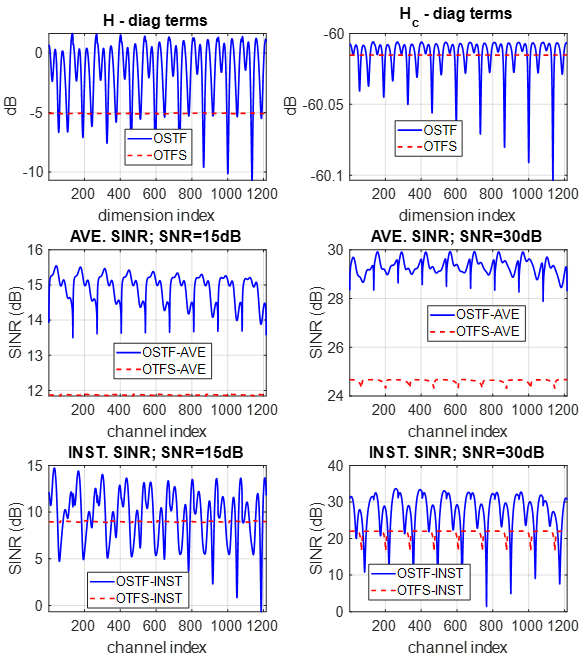}}
\vspace{-2mm}
\caption{\footnotesize{\sl Top: plots of the diagonal terms of $\bH$ and $\bH_c$ for OTSF and OSTF. Middle: plots of the {\em average} SINR in the different channels at the output of the MMSE filters, at two different operating $\snr$s. Bottom:  Plots for {\em instantaneous} $\sinr$ in the different channels.}}
\label{fig:diag_sinr}
\vspace{-6mm}
\end{figure} 
Fig.~\ref{fig:diag_sinr} plots the diagonal terms of $\bH$ and $\bH_{\rm c} = \bW^\dagger \bH$  in the top two panels for both systems. The middle two panels plot the {\em average} SINR in the MMSE outputs of the two systems at two different operating SNRs (15dB and 30dB), and the last two panels plot the {\em instantaneous} SINRs for a particular channel realization. We make a few important observations. First, the diagonal terms of $\bH_{\rm stf}$ show a lot more variation than those for $\bH_{\rm tfs}$.  This variation is significantly reduced after MMSE filtering as evident from the diagonal terms of $\bH_{\rm c,stf}$ and $\bH_{\rm c,tfs}$. Second, the {\em average} $\sinr$ at the output of the MMSE filter is uniformly higher for OSTF than OTFS.  On the other hand, the {\em instantaneous} $\sinr$ at the output of the MMSE filter is not uniformly higher in OSTF and exhibits significant fluctuations around the relatively uniform {\em instantaneous} $\sinr$ for OTFS.    These observations are consistent with a key property of  OTFS  noted in \cite{hadani:wcnc17}: all OTFS symbols experience identical gain in the diagonal elements of $\bH_{\rm tfs}$.  However, as evident from the plots of {\em average} $\sinr$ in (\ref{fig:diag_sinr}) this lack of fluctuation in the diagonal terms of $\bH_{\rm tfs}$ and $\bH_{\rm c,tfs}$ also leads to a loss in {\em average} $\sinr$ in OTFS compared to OSTF. 

\begin{wrapfigure}{l}{0.25\textwidth}
\vspace{-4mm}
\includegraphics[width=0.25\textwidth]{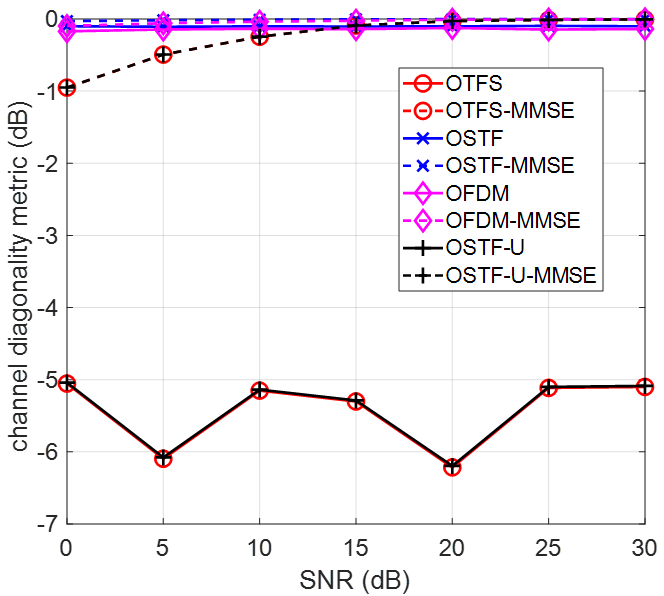}
\vspace{-6mm}
\caption{\footnotesize{\sl Plots of the diagonality metric for OTFS, OSTF, OFDM,  OSTF-U.}}
\label{fig:ch_metric}
\vspace{-4mm}
\end{wrapfigure}

Fig.~\ref{fig:ch_metric} plots the average diagonality metric, $E[\gamma(\bH)]$, for  
$\bH_{\rm tfs}$,  $\bH_{\rm c,tfs}$, $\bH_{\rm stf}$, and $\bH_{\rm c,stf}$. The OSTF metric  is about 
$5$dB larger than that for OTFS prior to MMSE filtering ($\bH$)  and the two  become comparable after MMSE filtering ($\bH_c$).    Note that $\gamma$ in nearly identifical for  OTFS and OSTF-U. Furthermore, $\gamma$ for OFDM is slightly lower than  OSTF.  

\begin{figure}[htb]
\vspace{-1mm}
\begin{tabular}{cc}
\hspace{-1mm}
\includegraphics[width=1.7in]{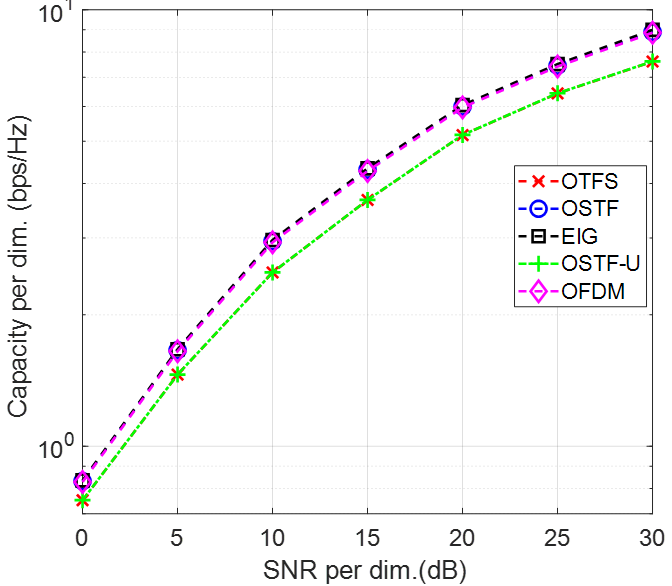}  & \hspace{-6mm} 
\includegraphics[width=1.7in]{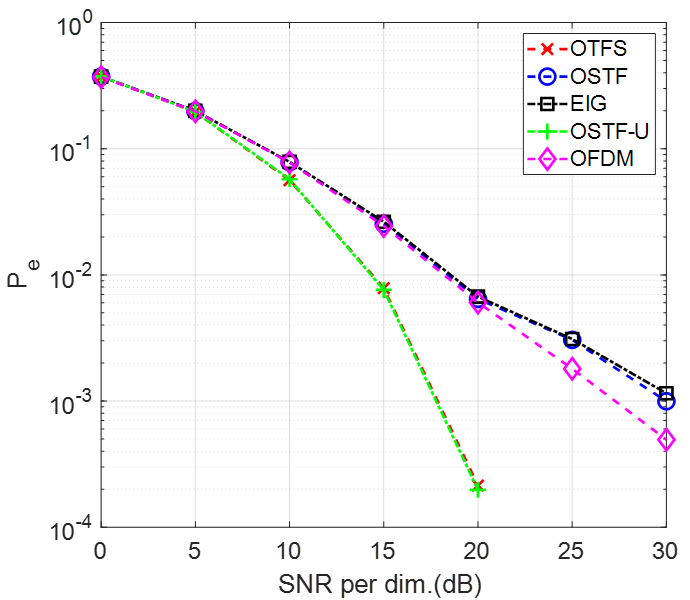} \\
\hspace{-1mm} {\footnotesize{(a)}} &  \hspace{-6mm}   {\footnotesize{(b)}}
\end{tabular}
\vspace{-2mm}
\caption{\footnotesize{\sl Plots for (a) capacity and (b) $P_e$ for  OTFS, OSTF, OSTF-U, and OFDM.}}
\label{fig:cap_pe}
\vspace{-1mm}
\end{figure}

Fig.~\ref{fig:cap_pe} compares the SINR-based capacity estimates and the $P_e$ for the two systems as a function of $\snr$ per dimension, averaged over $100$ channel and noise realizations. The plots show that the capacity of the OSTF system is nearly identical to the capacity of the eigen (EIG) system,  and higher than that for the OTFS system. The $P_e$ plots, on the other hand, show that the uncoded $P_e$ for OTFS is better than that for OSTF. This  is primarily related to the larger $\sinr$ fluctuations in OSTF compared to OTFS as shown in Fig.~\ref{fig:diag_sinr}.  Interestingly, the $P_e$ for OSTF is nearly identical to the EIG system.
The OFDM performance is comparable to OSTF, whereas the performance of OSTF-U is near identical to OTFS. The near-identical performance of  OTFS/OSTF-U suggests that any unitary pre-processing (and a corresponding post-processing  at the RX) suffices to smoothen out the channel fluctuations in the diagonal entries of $\bH$ and $\bH_c$ in OSTF. 
The comparable performance of OFDM to OSTF indicates that with full-CSI MMSE processing at the RX, even OFDM  can mitigate the interference induced by moderate Doppler.  This begs the question: {\em Do we need OSTF or OTFS in highly mobile environments, when OFDM with full-CSI MMSE performs nearly the same?} 
\begin{figure}[htb]
\vspace{-4mm}
\begin{tabular}{cc}
\hspace{-1mm}
\includegraphics[width=1.7in]{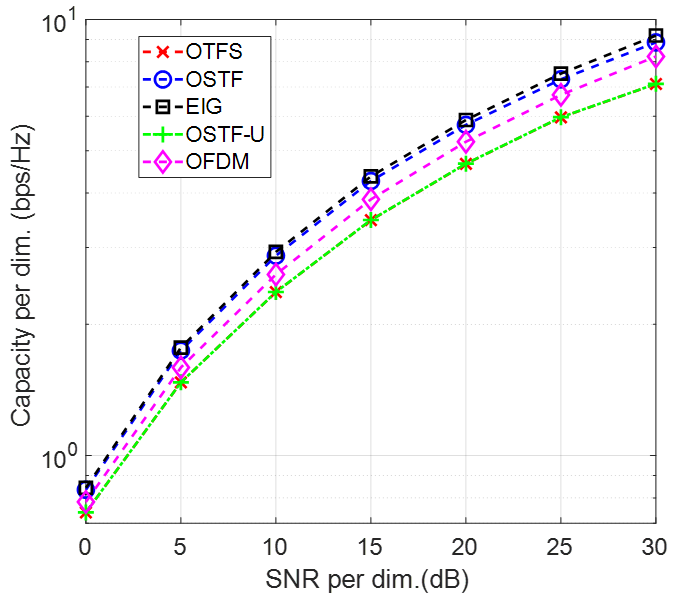}  & \hspace{-6mm} 
\includegraphics[width=1.7in]{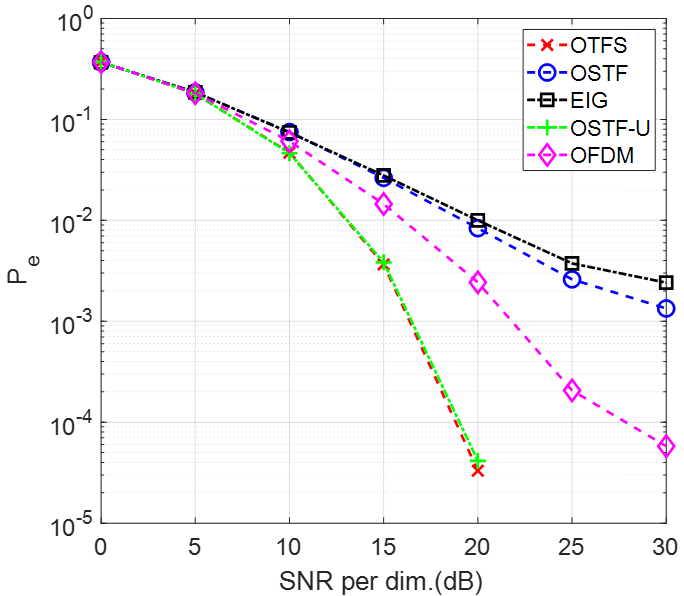} \\
\hspace{-2mm} {\footnotesize{(a)}} &  \hspace{-6mm}   {\footnotesize{(b)}}
\end{tabular}
\vspace{-2mm}
\caption{\footnotesize{\sl  Plots for the large channel spread scenario for (a) capacity and (b) $P_e$ for  OTFS, OSTF, OSTF-U, OFDM, and EIG using full CSI.}}
\label{fig:cap_pe_ls}
\vspace{-2mm}
\end{figure}

\begin{figure}[htb]
\vspace{-0mm}
\begin{tabular}{cc}
\hspace{-3mm}
\includegraphics[width=1.7in]{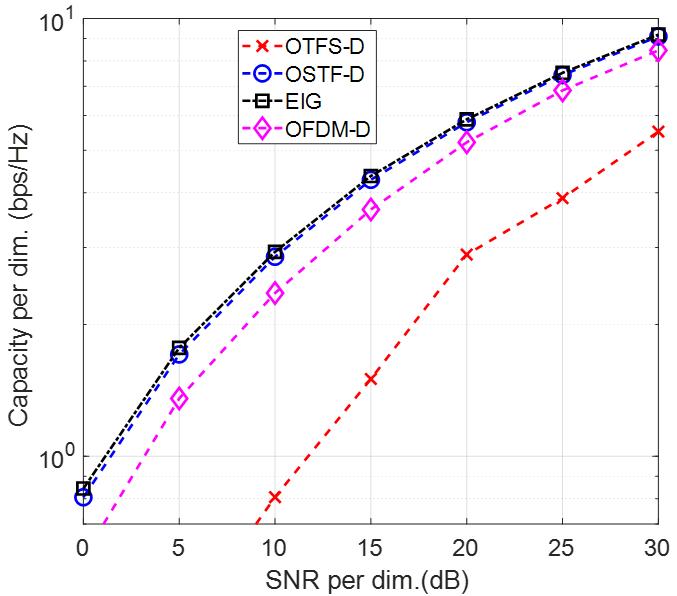}  & \hspace{-6mm} 
\includegraphics[width=1.8in]{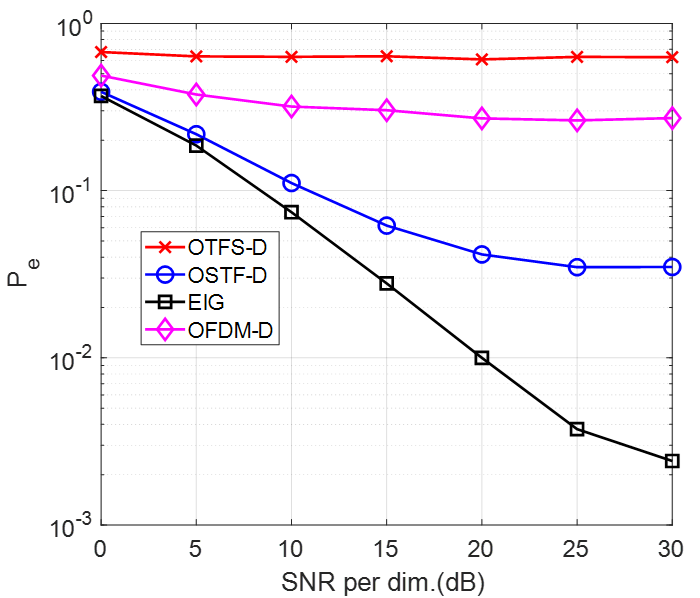} \\
\hspace{-1mm} {\footnotesize{(a)}} &  \hspace{-6mm}   {\footnotesize{(b)}}
\end{tabular}
\caption{\footnotesize{\sl Plots for the large channel spread scenario for (a) capacity, and (b) $P_e$, for the OTFS-D, OSTF-D, OFDM-D, and EIG systems.}}
\label{fig:cap_pe_d_ls}
\vspace{-4mm}
\end{figure}

To explore this question, we consider a propagation environment with larger delay and Doppler spreads to amplify the degradation in OFDM due to mobility: $\tau_{max} =700$ns and $\nu_{max} = 9.26$kHz (2500 kmph max. speed). In this case, $N_t=13$, $N_f = 93$ and $N = N_t N_f = 1209$.  
Fig.~\ref{fig:cap_pe_ls} plots the capacity and $P_e$ for the different systems with full CSI. Interestingly, OFDM has better $P_e$ than OSTF, due to less wild fluctuations in its diagonal entries, and OTFS/OSTF-U have the best $P_e$, as before.
We next consider system configurations in which only {\em diagonal} CSI  -  the diagonal entries of $\bH$ - is used for designing the MMSE filter $\bW$, motivated by the approximate eigen-property of OSTF. Note that $\bW$ also has a diagonal form in this case. Fig.~\ref{fig:cap_pe_d_ls} plots the capacity and $P_e$ for the different systems with only diagonal CSI. (EIG system uses full CSI and its performance is unchanged.)
Remarkably, the capacity of OSTF is near identical to EIG and is best, followed by OFDM and then OTFS (which incurs significant loss compared to Fig.~\ref{fig:cap_pe_ls}). The $P_e$ of OTFS is worst, OFDM is slightly better, OSTF is further better and EIG is the best. 

\vspace{-1mm}
\section{Concluding Remarks}
\label{sec:conc}
The framework developed in this paper explores the intimate relationship between OTFS and OSTF and also enables a unified comparison between OTFS, OSTF, OFDM and related schemes, e.g. \cite{zemen:pimrc18,fettweis:gcom18}, all of which assume an underlying ``time-frequency'' (OSTF) modulation. The opposing trends in capacity and $P_e$ for OSTF and OTFS warrant further investigation. The high-Doppler results in Figs.~\ref{fig:cap_pe_ls}-\ref{fig:cap_pe_d_ls} underscore the significance of using appropriately designed OSTF basis waveforms for doubly dispersive channels and the lower level of CSI needed for OSTF. The results suggest new possibilities between transmission on a localized OSTF basis and spreading the symbols over all basis functions (OTFS); e.g, $\bU_{\rm sfft}  = \bU_{N_f}^{\dagger *} \otimes \bU_{N_t}^\dagger $ has a kronecker structure that could be exploited for designing the precoding matrix akin to structured linear dispersion codes \cite{sayeed:04vtc}. The use of coding may also close the $P_e$ gap between uncoded OSTF and OTFS systems.

\vspace{-1mm}
\bibliographystyle{IEEEtran}
\bibliography{otfs_vs_stf_gcom21_final.bbl}

\end{document}